\documentclass[runningheads]{llncs}
\usepackage[T1]{fontenc}
\usepackage{multirow}
\usepackage{graphicx}
\usepackage{hyperref}
\usepackage{caption}
\usepackage{xcolor}
\usepackage{subfig}
\begin{document}
\title{The Road Less Traveled: Investigating Robustness and Explainability in CNN Malware Detection}
%
%
\author{Matteo Brosolo\inst{1}\orcidID{0000-1111-2222-3333} \and
Vinod P\inst{2,3}\orcidID{1111-2222-3333-4444} \and
Mauro Conti\inst{3}\orcidID{2222--3333-4444-5555}}
\authorrunning{M. Brosolo et al.}
%
\institute{University of Padova, Italy 
\email{matteo.brosolo@unipd.it}\\
}
\maketitle              
\begin{abstract}
   Machine learning has become a key tool in cybersecurity, improving both attack strategies and defense mechanisms. Deep learning models, particularly Convolutional Neural Networks (CNNs), have demonstrated high accuracy in detecting malware images generated from binary data. However, the decision-making process of these black-box models remains difficult to interpret. This study addresses this challenge by integrating quantitative analysis with explainability tools such as Occlusion Maps, HiResCAM, and SHAP to better understand CNN behavior in malware classification. We further demonstrate that obfuscation techniques can reduce model accuracy by up to 50\%, and propose a mitigation strategy to enhance robustness. Additionally, we analyze heatmaps from multiple tests and outline a methodology for identification of artifacts, aiding researchers in conducting detailed manual investigations. This work contributes to improving the interpretability and resilience of deep learning-based intrusion detection systems.

\end{abstract}
\section{Introduction}
Technology has rapidly evolved over the past decade, providing individuals and corporations with advanced tools tailored to their needs. However, this growth, along with widespread internet accessibility, has also created opportunities for cyber threats. The rapid increase in malware presents a major challenge for cybersecurity experts\footnote{\href{https://portal.av-atlas.org/malware}{Av-Test, Accessed 2023-5-15}}. Its relentless growth shows no signs of abating, with a continuous surge in reported cases across both security and non-cybersecurity corporations, including government agencies. 
The mounting pressure on cybersecurity research has encouraged the creation of novel detection and classification techniques aimed at countering the escalating sophistication of modern malware~\cite{vignau2021theevolution,gibertsurvey}.
Modern antivirus solutions rely on machine learning, especially deep learning techniques, to enhance threat detection and prevention. Beyond traditional signature-based detection, these advanced algorithms offer a novel approach to analyzing threats. One promising technique involves visualization, where a binary file is converted into an image to represent suspected malware. This approach enables the use of well-established computer vision tools and models for malware analysis~\cite{gopinath2023acomprehensive}. Since Nataraj et al.~\cite{nataraj2011malware} introduced malware visualization in 2011, researchers have developed various techniques to improve visual classifiers for classifying malware families, achieving state-of-the-art results on benchmark datasets~\cite{kumar2024imcnn}. However, many of these models have been adopted without a deep understanding of their decision-making processes. Most high-performing visual models rely on black-box architectures, such as Convolutional Neural Networks (CNNs), which offer strong predictive performance but lack interpretability. This lack of transparency poses challenges, especially when models fail or require modifications for new threats. When analyzing advanced malware, equipping human analysts with supportive tools—such as neural networks—enhances detection and response. A more interpretable system fosters smoother collaboration between humans and AI~\cite{ras2021explainable,aonzo2023humans}. Beyond practical concerns, governments worldwide are increasingly prioritizing machine learning explainability to ensure fairness in decision-making~\cite{voigt2017theeu}.

\par In this study, we aim to assess the robustness of a state-of-the-art CNN classifier by evaluating its performance against a newly constructed dataset designed to test specific scenarios. We investigate how CNN classifiers respond to common adversarial tactics employed by malware developers, including obfuscation and packing. At each stage, we leverage interpretability tools, such as Class Activation Maps~(CAMs)~\cite{zhou2015learning}, SHAP~\cite{lundberg2017unified} values, and occlusion maps~\cite{occlusion} to explain the reasons behind the model’s success or failure. Our findings demonstrate how these eXplainable Artificial Intelligence~(XAI)~\cite{ras2021explainable} tools can benefit defenders in developing more resilient defenses and improving classifier performance.

This study aims to answer the following research questions:  
\begin{itemize}
\item \textbf{RO1}: How robust is a malware visualization CNN model when evaluated on a modern dataset with diverse malware samples?  
\item \textbf{RO2}: Can deep learning-based visualization models effectively detect obfuscated or packed malware, or do these techniques degrade their performance? 
\item \textbf{RO3}: How effectively do black-box interpretability tools (SHAP, HiResCAM, and occlusion) explain CNN predictions, and under what conditions do they produce conflicting interpretations of the model’s decision-making process?
\end{itemize}
The contributions we made with this research are:
\begin{enumerate}
    \item We introduced a new dataset designed to better represent emerging malware and its various forms as observed in the wild.
    \item To assess the robustness of state-of-the-art CNN models, we conducted experiments using different dataset modifications. Specifically, we performed partial training and benchmarking on a standard dataset. Additionally, we proposed a simple dataset enhancement technique that improves accuracy on obfuscated malware by up to 30\% without adding new samples.
    \item To gain insights into the model’s classification process, we applied three different Explainable AI techniques and conducted an in-depth analysis of their results. This allowed us to identify varying levels of understanding within the CNN, based on the regions it highlights in the input samples.
\end{enumerate}
\section{Dataset}
\label{sec:dataset}
When talking about malware classification on the Windows platform, which is by far the historically most targeted, two datasets emerge as the de facto standard used by researchers when benchmarking their models. This two datasets are called MalImg~\cite{nataraj2011malware} and Big2015~\cite{ronen2018microsoft}.

We evaluate our model's performance on benchmark datasets to demonstrate that it represents a state-of-the-art classifier. The primary difference between our model and the highest-performing visual malware classifiers lies in the additional techniques they employ, such as embedding extra information in images before training or using ensemble methods. For example, some techniques enhance classification performance by embedding metadata, such as file entropy. Others modify the visualization by applying transformations like color encoding, feature stacking, or multi-channel representations to highlight key malware characteristics. These approaches improve detection but are not the focus of our research, as they still rely on a well-performing CNN, similar to the one we developed. Analyzing such modifications would reduce generalizability and offer limited new insights to the field.

After checking other datasets that still did not provided what needed, we decided to develop a dataset of our own. The results of our effort is VXZoo, which has been summarized in Table~\ref{tab:table-ogds}. The fundamental principles guiding the development of this malware dataset included:
\begin{enumerate}
    \item \textbf{Incorporating new malware}:  As malware continuously evolves, we designed a dataset that represents current threats. MalImg and Big2015 are becoming outdated, and testing on them may no longer accurately reflect the behavior of modern malware variants;
    
    \item \textbf{Year-wise identification}: To analyze how malware evolves, we included samples from different years whenever possible. This approach helps researchers evaluate classifier performance against malware that has mutated over time;
    
    \item \textbf{Maintaining imbalanced characteristics}: Emulating the natural variation observed in the wild, where malware classes are not evenly distributed, we preserved the imbalanced nature of sample distribution. This decision aligns with the reality of cybersecurity threats, where certain malware classes are widespread, while others are more niche;
    
    \item \textbf{Ensuring variety in malware types}: We designed the classifier to detect malware broadly rather than specializing in a specific type. While incorporating diverse malware samples, we ensured the model could distinguish between classes without overemphasizing differences between malware types;
    
    \item \textbf{No distinction between obfuscated and non-obfuscated malware}: Instead of grouping all obfuscated malware into a single category, we treated obfuscated and non-obfuscated samples the same. Our goal was to ensure classifiers accurately identify malware classes, regardless of obfuscation;
    
    \item \textbf{Increased sample size and intermediate class count}: In our endeavor to enhance the dataset’s robustness, we aimed for a larger sample size than both Big2015 and MalImg. Additionally, we settled on a number of classes that fell between the extremes, striking a balance that allowed for a comprehensive evaluation of classifier performance.
\end{enumerate}

To compile the dataset, we obtained malware binaries from VirusShare\footnote{\href{https://virusshare.com}{VirusShare}, Accessed: 2023-5-15}, stored in ZIP archives accessed via torrents. We selected ZIP files uploaded between 2015 and 2022, choosing a representative sample for each year. We used the Kaspersky\footnote{\href{https://www.kaspersky.it/}{Kaspersky}, Accessed: 2023-5-15} naming convention from the VX-Underground\footnote{\href{https://www.vx-underground.org/}{VX-Underground}, Accessed: 2023-5-15} repository for labeling. Using these labels, we identified and removed non-PE (Portable Executable) files from the archives. We then focused on 15 malware families that appeared consistently over the eight-year period. The dataset reflects the evolution of malware over time, with some families showing sporadic spikes in specific years, while others maintained steady or clustered occurrences across consecutive years.

We transformed the malware in grayscale image using the standard technique used in Nataraj et al.~\cite{nataraj2011malware} and adapted for Big2015 in the form of the B2IMG algorithm used in Tekerek et al.~\cite{tekerek2022anovel}. 
The data in the sample file are read line by line, and each line is split into an array. The hexadecimal values in the array are converted into decimal numbers ranging from 0 to 255. These decimal values are used to generate pixels for grayscale images. A matrix is created based on the calculated aspect ratio with all initial values set to zero. The decimal values, ranging from 0 to 255, are then inserted into the matrix. Finally, images of each malware were obtained using the 2D matrix.

\begin{figure}
\centering
\subfloat[Visual Basic]{\includegraphics[width=0.19\textwidth]{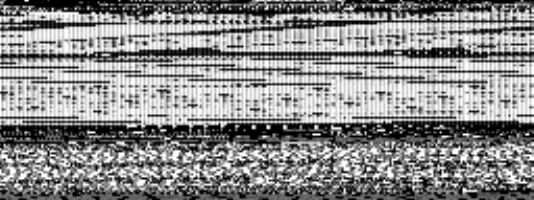}\label{fig:Ex_Im}}\quad
\subfloat[Injector]{\includegraphics[width=0.19\textwidth]{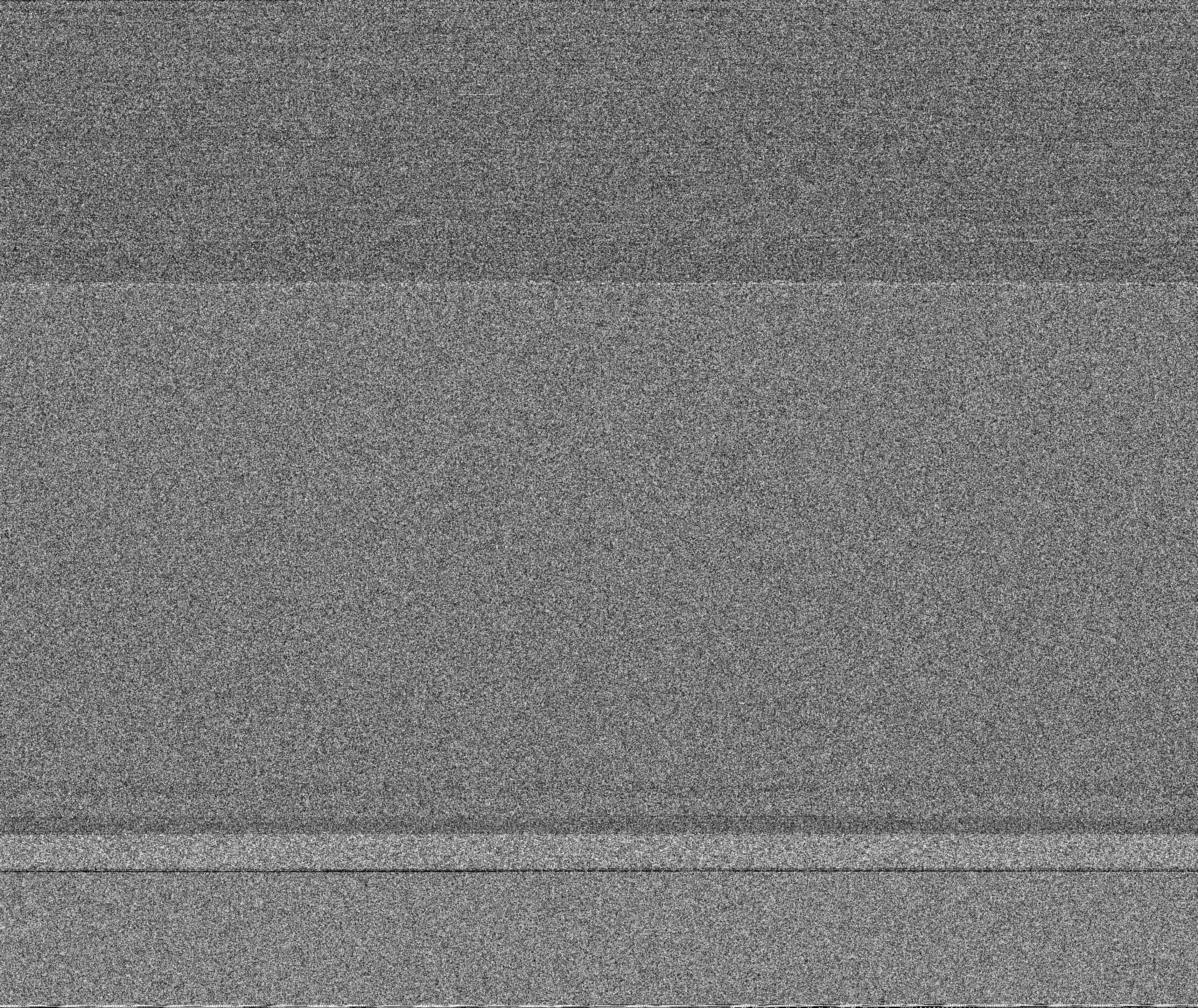}\label{fig:Ex_Im2}}\quad
\subfloat[Flooder]{\includegraphics[width=0.19\textwidth]{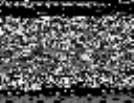}\label{fig:Ex_Im4}}
\caption{Examples of different malware families' samples.}\label{fig:examples_general}
\end{figure}

We provide a brief overview of the selected malware families and their characteristics. This discussion not only highlights the types of malware the model can detect but also explains how each class’s unique traits appear in visual representations. For instance, flooders are typically smaller in size, resulting in more compact images. Additionally, different malware techniques often exhibit distinct graphical patterns. Figure~\ref{fig:examples_general} illustrates examples of these malware structures through cumulative images.

\begin{enumerate}
\item \textbf{Alman} (or Almanahe) is a common family of viruses that attack the Windows platform. They genearlly use the APC\footnote{Asynchronous Procedure Call} technique to inject their code in all the filesystem executables.

\item \textbf{AntiFW} Kaspersky provides the heuristic AntiFW to identify Trojan horse malware. Although not a distinct malware family, AntiFW serves as a ``catch-all" category for files exhibiting Trojan characteristics that do not fit into other classifications, regardless of design variations.

\item \textbf{Debris} is a class of worm malware that usually spreads via email attachments.

\item \textbf{Vtflooder} is a special type of malware that attacks platforms like VirusTotal in order to ``flood" them and cause a denial of service.

\item \textbf{Elkern} is a very old class of parasitic viruses that even though uses malicious spreading techniques, it doesn't do anything overly aggressive to the host platform. While our dataset primarily consists of modern malware samples across various classes, we have intentionally included one class containing older samples for a variety of reasons: robustness, legacy compatibility and comprehensive benchmarking.

\item \textbf{Expiro}: Is a stealer, having 
capabilities to install browser extensions, change security behaviour/settings on the infected system, and steal information

\item \textbf{Inject}: It's another general class containing malicious programs that inject their code into the address space of programs running on the infected computer, such as system processes or programs that have access to the Internet. 

\item \textbf{Lamer}: Kaspersky uses the Lamer classification to group malware based on specific tactics and techniques, particularly those related to defense evasion, privilege escalation, and persistence.

\item \textbf{Nimnul}: also known as Ramnit, is a very famous banking trojan. It was used as a class in the Big2015 dataset as well.

\item \textbf{Parite}: is a memory-resident polymorphic virus that infects executable files with EXE and SCR extensions. It installs a dropper file onto the system.

\item \textbf{Sality}:  initially developed in 2003, is an advanced malware that has evolved over the years. Developers have continuously enhanced it by adding features like rootkit and backdoor functionality, making it a persistent and relevant threat despite its age.

\item \textbf{VB}: is a ``catchall" class for all the VisualBasic executables that have been identified containing malicious code.

\item \textbf{Virut}: is a botnet malware family which has initially been observed in 2006. Traditionally, it spreads as a file-infecting virus, and has monetized pay-per-install schemes and information theft. It has notably been targeted in 2013 by the NASK/CERT polish cybercrime agency. Even if the sinkholing effort has been largely successful, samples are still found in the wild.

\item \textbf{Wabot: }is an IRC worm written in Delphi that drops multiple files into the \textit{system32} directory, and creates a text file containing ASCII art of the word \textit{marijuana} in the system root. It connects to an IRC server, joins the channel \textit{\#HelloThere}, and awaits backdoor commands.

\item \textbf{WBNA}: is a worm that infiltrates systems either as a dropped file from other malware or as a hidden download from malicious websites.
\end{enumerate}

\par We use Kaspersky's classification to categorize malware based on shared characteristics rather than specific families defined by analysts. This approach provides a broader classification that groups malware by generic techniques instead of strict family names. This method is more realistic because, in the real world, some malware may not yet belong to a known family, but may still exhibit malicious behavior. A broader grouping (coarse labeling) helps classify these threats effectively. Additionally, this approach challenges the classifier, especially a CNN, by requiring it to distinguish between similar malware types (e.g., worms), while also differentiating between well-defined malware families and loosely grouped categories.


\section{Methodology}

This section outlines the methodology used to test model robustness and extract explainability maps. We explain the difference between interpretability methods and describe how these maps enhance classification performance and provide qualitative insights into predictions.

\subsection{Benchmark CNN Classifier}
To evaluate the performance of the malware visualization approach, we designed a state-of-the-art model that reflects commonly used architectures in the field. Our design draws inspiration from Gibert et al.~\cite{gibert2019using} and the VGG16 architecture~\cite{simonyan2014very}. VGG16 is a widely used neural network in computer vision and has been extensively adapted for image-based malware analysis, consistently delivering competitive results~\cite{vasan2020imcec,vasan2020imcfn}. Our CNN architecture consists of three main blocks, each containing a convolutional layer, ReLU activation, pooling, dropout, and normalization. The output is then flattened and processed through two dense layers, followed by a softmax function for malware family classification. We excluded models such as DenseNets, EfficientNets, and Inception networks because their complexity exceeds our study's requirements. Although these models perform well on malware datasets~\cite{tekerek2022anovel,chaganti2022image}, their specialized architectures, such as Dense blocks in DenseNets or uniform scaling in EfficientNets, might limit the generalizability of our explainability insights. 
To further refine our model, we applied hyperparameter tuning using the Keras Tuner~\cite{kerastuner} and optimized eight hyperparameters with the HyperBand technique. The tuning results are presented in Table~\ref{tab:hyperparams}.

\begin{table*}[ht]
\centering
\caption{Hyperparameters of Classification Model}
\label{tab:hyperparams}
\begin{tabular}{|l|l|}\hline
\textbf{Hyperparameter} & \textbf{Value After Tuning} \\ \hline
Kernel Size First Conv  & 5                           \\
Lambda Normalization    & True                        \\
First Dense Layer Size  & 1024                        \\
Second Dense Layer Size & 256                         \\
Dropout Conv            & 0.1                         \\
Dropout Dense           & 0.3                         \\
Learning Rate           & 0.0003                      \\
Momentum                & 0.95                       \\ \hline
\end{tabular}
\end{table*}

\par We evaluated our model on two benchmark datasets, MalImg and Big2015, to assess its performance. Without dataset-specific tuning or advanced techniques like transfer learning or ensemble methods, the model still demonstrated strong classification capabilities. These results confirm that our implemented model is suitable for further study and explainability analysis in this research.

We tested the VXZoo dataset with a partial training set to evaluate the model's effectiveness and determine the threshold at which its performance significantly improves. The experiments involved training on 20\%, 40\%, 60\%, and 80\% of the dataset. Additionally, we assessed the model's robustness against adversarial samples using common real-world evasion techniques. Specifically, we applied packing and metamorphism, utilizing UPX\footnote{\href{https://github.com/upx/upx}{UPX}, Accessed: 2023-5-15}~\cite{upx}, one of the most widely used packers and the defacto standard for malware authors~\cite{upx_common}.

The UPX packer is an open-source tool that, while not inherently malicious, is frequently used by hackers to obscure malware content. Due to the inherent characteristics of certain malware classes and individual specimens, we could only pack some samples.  

For metamorphism, we utilized a readily available metamorphic engine that applies basic transformation techniques to each sample. We selected pymetangine\footnote{\href{https://github.com/scmanjarrez/pymetangine}{pymetangine, Accessed 2023-5-15}} because it introduces simple modifications to the code, allowing us to assess whether even basic metamorphic techniques impact the classification of a visual model.  

We packed every sample in the dataset and applied metamorphic transformations to each sample at least three times. Samples and families were selected randomly, meaning that, as seen in real-world scenarios, not all files could be packed or morphed. This variability is significant because a malware family's ability to be packed or morphed affects its detectability. Some families are more susceptible to these transformations, while others cannot be modified due to intrinsic limitations.

Table~\ref{tab:dataset_morph_pack} shows the conversion rate of each class based on the number of morphed and packed samples. Notably, if a sample can be morphed once, it remains morphable indefinitely. This study evaluates the performance of a representative CNN on a new dataset and investigates the factors influencing its behavior. By analyzing explanation maps, we aim to identify artifacts in malware images that reveal distinctive binary features, enhancing classifier performance and generating new malware signatures. Improving the interpretability of black-box models also helps researchers understand model decisions, making them more transparent and actionable for malware analysis.

After analyzing the raw metrics that compare the performance of modified and original samples, we applied explainability techniques. To ensure a diverse evaluation, we selected three methods with different explanation strategies: HiResCAM, SHAP, and occlusion maps.

\begin{itemize}
  \item \textbf{HiResCAM~\cite{hirescam}}: HiResCAM, a variant of GradCAM, has demonstrated strong performance in explainability tasks. This Class Activation Mapping (CAM) technique identifies image regions that contribute most to the predicted class. It generates activation heatmaps by combining the last convolutional feature map of the CNN with the output layer's weights, highlighting areas that strongly influence the model's decision. The heatmap reveals the features that maximize the CNN's confidence in its prediction but does not indicate how the model responds to perturbations or removed features. 
  
    \item \textbf{SHAP}~\cite{shap}: SHAP is a game-theory-based approach that assigns a Shapley value to each pixel or group of pixels, which represents the contribution of that feature to the CNN’s prediction. It works by creating a large number of perturbed versions of the original image and then measuring how the prediction changes in each case. Based on this, SHAP calculates the average contribution of each group of pixel to the final prediction. Unlike CAM, which focuses on activations, SHAP quantifies the contribution of each areas in a way that satisfies theoretical fairness properties.

    \item \textbf{Occlusion Maps}~\cite{occlusion}: This technique evaluates a CNN`s sensitivity by systematically masking different parts of an image and recording changes in its predictions. The resulting heatmap highlights the most essential regions for classification by identifying areas where occlusion causes the greatest prediction change. Unlike HiResCAM, which shows where the model focuses, occlusion maps assess the model’s reliance on specific regions by testing the impact of information removal.
    
\end{itemize}
Each of these three techniques employs different principles to identify the most and least important areas of an image from the CNN’s perspective. Beyond analyzing their individual insights, we examine where they align and diverge, providing explanations for these differences. When the techniques consistently highlight the importance of a specific region, we conduct further testing on relevant samples to validate the model’s ability not only to predict but also to explain its decisions. In particular, we analyzed how the interpretations altered between the original dataset and its modified version with packing and metamorphism. In addition to individual sample heatmaps, we introduced cumulative heatmaps to gain a broader understanding of how the CNN interprets an entire malware class.


Previous research, as well as our findings, highlight the importance of the PE header as a key feature for malware classification~\cite{peheader1,peheader2}.  To conclude the quantitative study, we incorporated morphed and packed samples into the training set and measured accuracy changes when testing on both the original and obfuscated datasets. Finally, by comparing the explainability heatmaps, we demonstrated that these tools not only reveal the CNN's classification reasoning but also help researchers identify artifacts in PE files that could be used in future signature-based classifiers.

\begin{table*}[ht]
    \centering
    \caption{VX-Zoo dataset composition. 
        \label{tab:table-ogds}}
    \begin{tabular}{|l|l|l|l|l|l|l|l|l|l|}
    \hline
    \textbf{Family} & \textbf{Total} &\textbf{2015} & \textbf{2016}&\textbf{2017} &\textbf{2018} &\textbf{2019} &\textbf{2020} &\textbf{2021} &\textbf{2022} \\ \hline
    Alman          & 225    & 218 & 3 & 0 &0 &2 &2 &1 &0   \\ 
    AntiFW         & 5777 & 2784 & 1419 & 1574 & 0 & 0 & 0 & 0 & 0     \\ 
    Debris         & 208   & 208 & 0 & 0 & 0 & 0 & 0 & 0 & 0 \\ 
    Elkern         & 219   & 219 & 0 & 0 & 0 & 0& 0& 0&0  \\ 
    Expiro         & 326  & 325 & 0 & 0 &0 &0 & 0& 1 &      38  \\ 
    Inject         & 275 & 0 & 0 & 275 & 0& 0&0 &0 &0    \\ 
    Lamer          & 748   & 16 & 0 & 0 & 17 & 27 & 636 & 44 & 14  \\
    Nimul          & 3043   & 37  & 0 & 0 & 4& 2& 7& 2984 & 10  \\ 
    Parite         & 380   & 361 & 7 & 1 & 1 & 7  &1 &2 &0     \\    
    Sality         & 851 & 809 & 7 & 1 &10 &15 &11 &2 &0     \\ 
    VB             & 130 & 0 & 0 & 0 &0 & 0& 120&4 & 6 \\ 
    Virut          & 1609 & 849 & 11 & 5 &7 &15 &647 &39 &39  \\ 
    Vtflooder      & 465 & 465 & 0  & 0&0 &0 &0 &0 &0    \\ 
    Wabot          & 442     & 60 & 0 & 0 &0 &0 &363 &4 &  15  \\ 
    WBNA           & 907     & 885 & 0 & 0 &2 &1 &1 &12 &7    \\ \hline

    \end{tabular}

\end{table*}

\section{Experimental Study}
This section presents the results of our pipeline, which tests model robustness and derives insights using explainability techniques. First, we introduce the benchmark and baseline classification task, then generate explainability heatmaps to correlate visual patterns with quantitative outcomes. In total, we conducted the following experiments:
\begin{itemize}
    \item Training and benchmark of the CNN on standard datasets (Sec.\ref{sec:benchmark})
    \item Progressive training to test robustness (Sec.\ref{sec:progressive}).
    \item Test on morphed and packed datasets (Sec.\ref{sec:morphpack}).
    \item Test on normal dataset, morphed and packed datasets with the enhancement technique (Sec.\ref{sec:morphpack}).
    \item Extraction of XAI heatmaps with HiResCAM, Occlusion Maps and SHAP and their complete analysis (Sec.\ref{sec:xai}).
\end{itemize}

\begin{table}[]
\centering
\caption{Number of samples that have been converted from normal to morphed or packed. }
\label{tab:dataset_morph_pack}
\begin{tabular}{|l|l|l|}\hline
\textbf{} & \textbf{Morph}   & \textbf{Pack}\\ \hline
\textbf{Percentage}                  & 83.30\%          & 25.18\%          \\
\textbf{Empty classes}                  & 2          & 4          \\ \hline
\end{tabular}

\end{table}
\subsection{Benchmark}
\label{sec:benchmark}
To initiate our study, we partitioned the two selected benchmark datasets, MalImg and Big2015, into training and test sets using an 80/20 ratio. We allocated 15\% of the training set for validation purposes. The results we obtained are consistent with current state-of-the-art performance, allowing us to proceed with subsequent phases of our research. Specifically, we achieved an accuracy of 98\% for MalImg and 95\% for Big2015. F1 score were respectively 97\% and 94\%. These results are comparable to the best-performing visual classifiers in the field that utilize similar experimental conditions as our CNN without incorporating additional improvements to the model like dataset augmentation or transfer learning. For context, top-performing models report accuracies around 99\%~\cite{kumar2024imcnn,tekerek2022anovel,vasan2020imcec} for both MalImg and Big2015.




\begin{figure}[h!]
    \centering
    \includegraphics[scale=0.4]{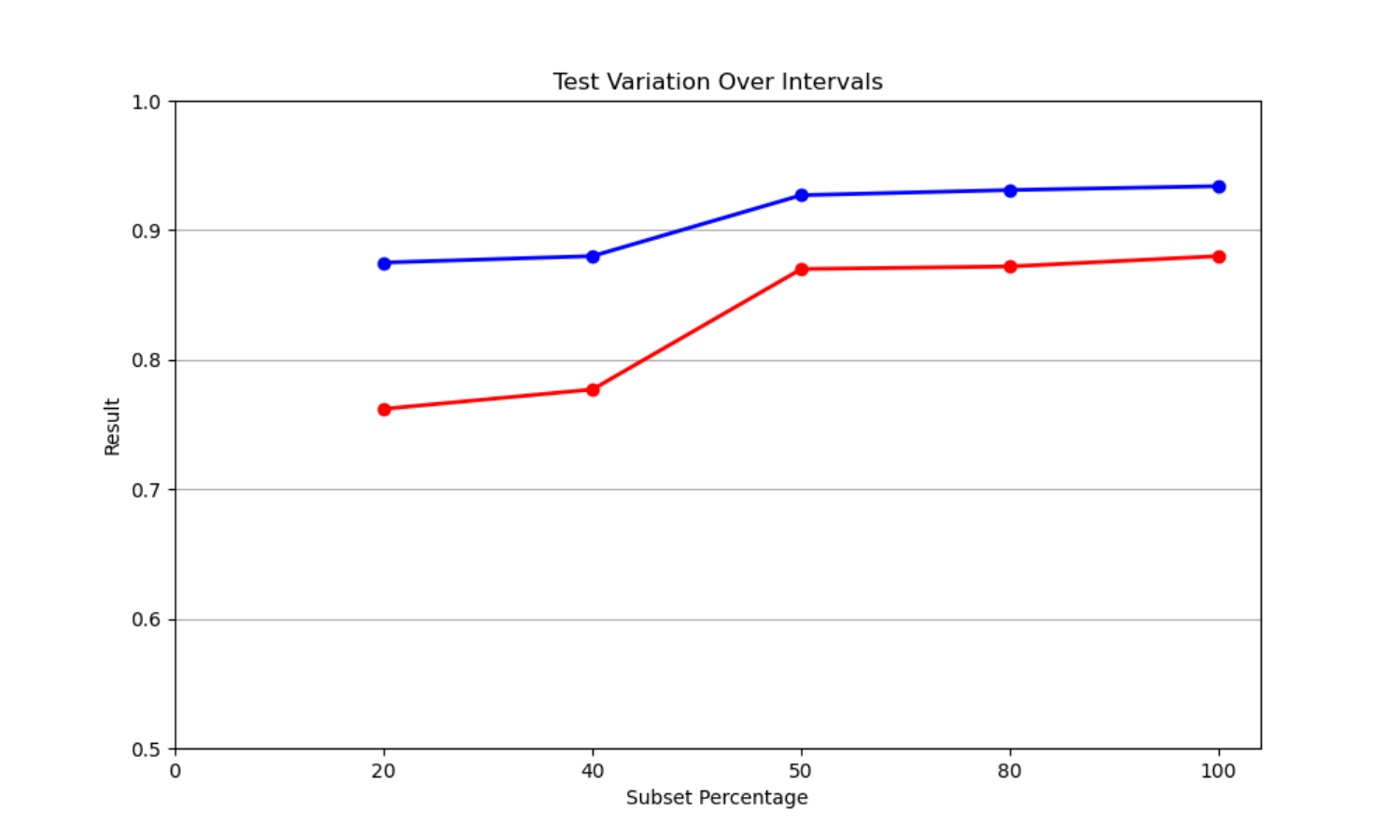}
    \caption{Evolution of partial training set evaluation.}
    \label{fig:subset_perc}
\end{figure}

\begin{figure}
    \centering
    \includegraphics[scale=0.26]{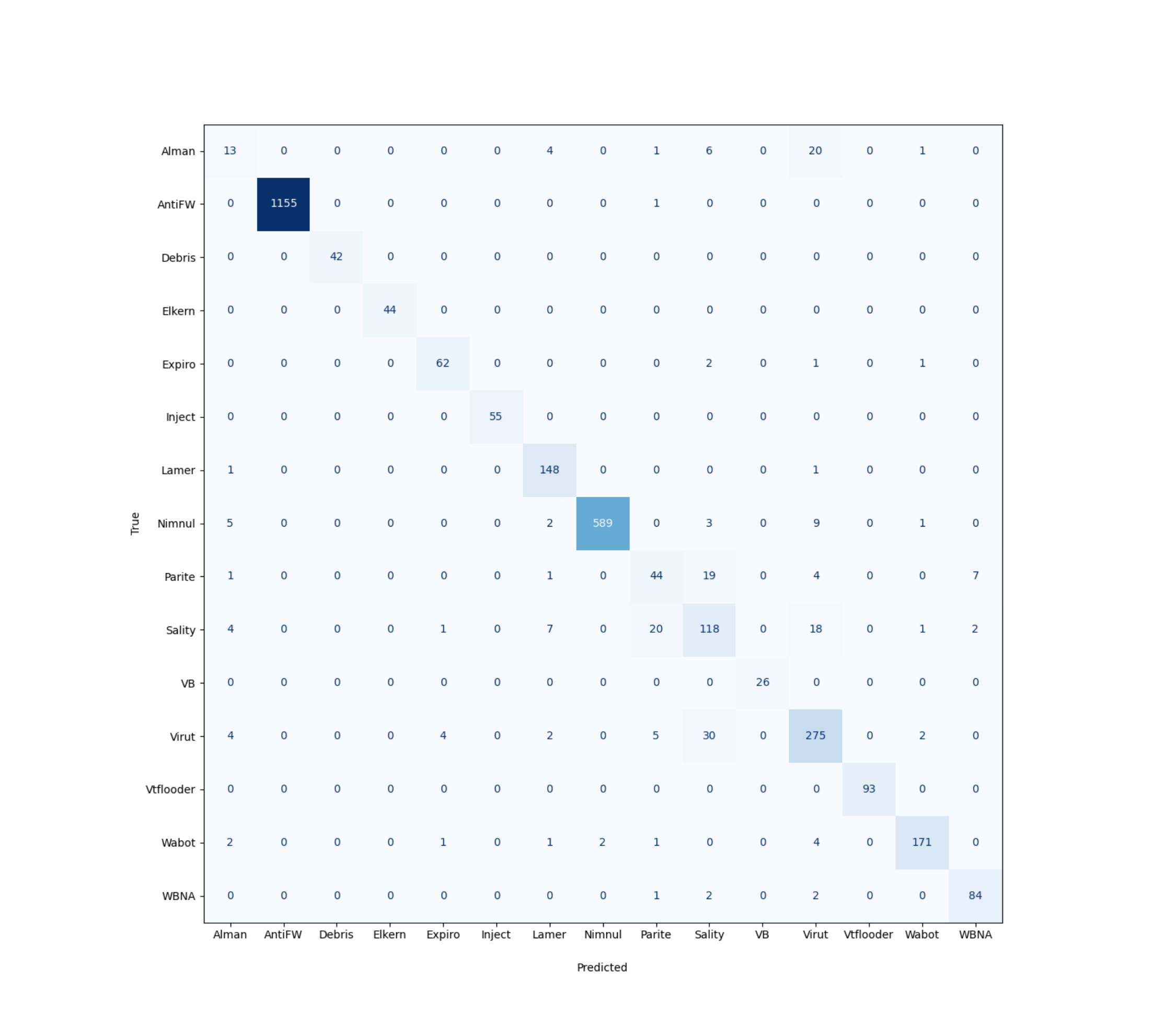}
    \caption{Confusion matrix for the base test set and base training set.}
    \label{fig:conf_matrix1}
\end{figure}

\subsection{Progressive Training}
\label{sec:progressive}
We subsequently conducted experiments on our VXZoo dataset using a progressive training approach. Testing a CNN model with varying percentages of the training set helps assess its data efficiency and learning behavior. By evaluating performance at different training set sizes, we can determine how much data is necessary for optimal learning. We selected subsets of 20\%, 40\%, 60\%, 80\%, and 100\% of the data to observe the evolution of model accuracy. The results, illustrated in Fig.~\ref{fig:subset_perc}, demonstrate the model's data dependency and reveal a plateau in accuracy at approximately 80\% of the training set, indicating that 12.484 overall samples were sufficient for our training set.
 The fact that the model performs well even at 20\% suggests strong generalization, while the improvement at 50\% indicates that additional data still provides benefits. This analysis helps in understanding the trade-off between dataset size and performance, guiding decisions on data collection and model scalability. Despite the marginal improvements beyond the 80\% threshold, we opted to proceed with subsequent experiments using the complete training set, as it yielded the highest performance, albeit by a small margin.

\subsection{Morph and Pack}
\label{sec:morphpack}
\begin{figure}
    \centering
    \includegraphics[scale=0.26]{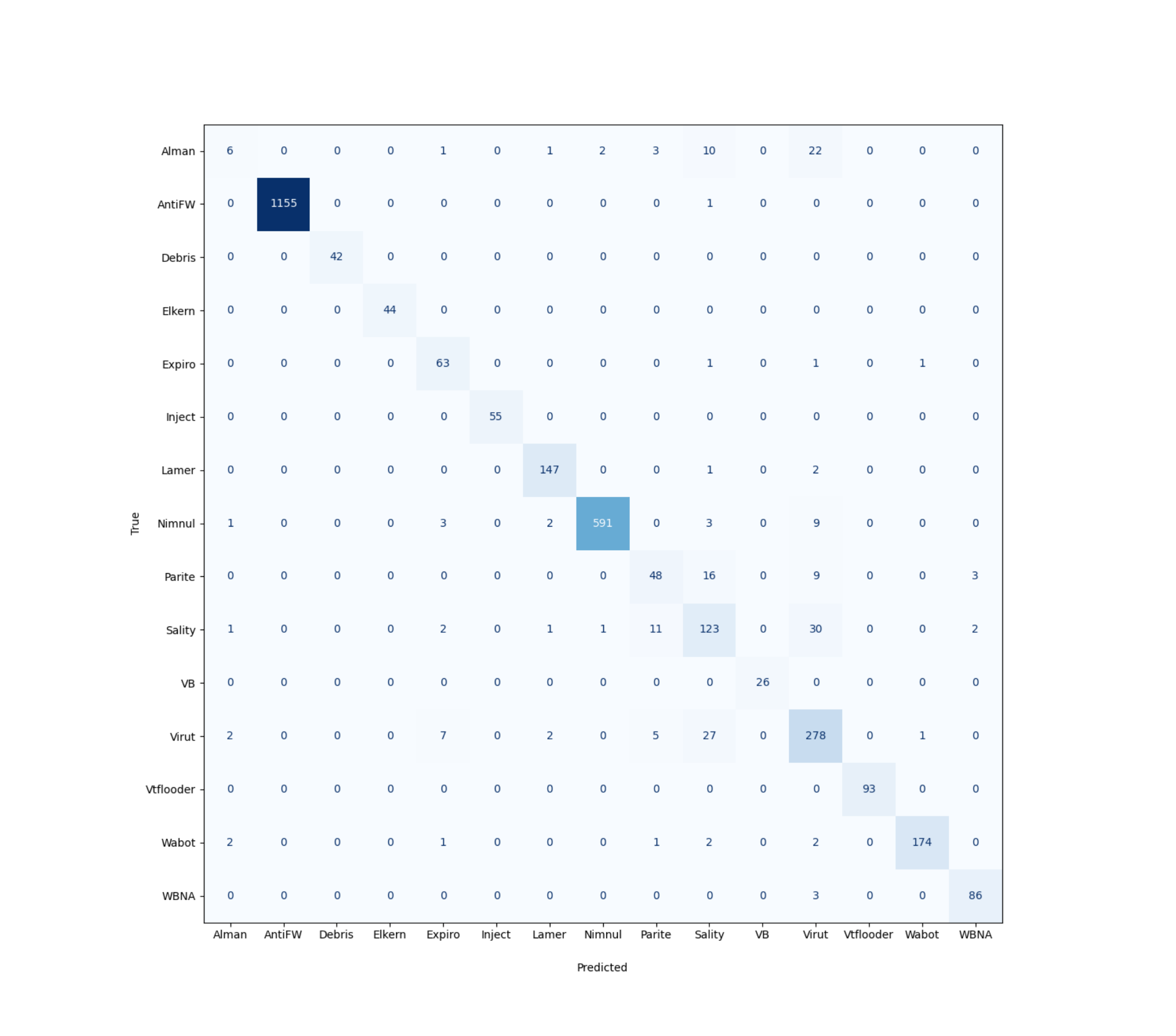}
    \caption{Confusion matrix for the base test set and enhanced training set.}
    \label{fig:conf_matrix2}
\end{figure}

In this phase, we introduced adversarial obfuscations commonly employed by malware authors, specifically packing (using UPX Packer) and morphing (using \textit{pymetangine}). Initially, we applied morphing to each sample that did not give errors in three stages. However, our preliminary tests revealed no significant variation in model accuracy between samples to which morphing is applied recursively once, twice, or three times. Consequently, we proceeded with testing only on samples morphed a single time. The results are presented in Table~\ref{tab:results_base}, Fig.~\ref{fig:conf_matrix1} and Fig.~\ref{fig:conf_matrix2}.

\begin{table}[]
\centering
\caption{Results obtained on the testing sets with enhanced and not enhanced datasets.}
\label{tab:results_base}
\begin{tabular}{|l|l|l|l|l|l|l|}\hline
\multirow{2}{*}{\textbf{Training Set}} & \multicolumn{2}{l|}{\textbf{Base}} & \multicolumn{2}{l|}{\textbf{Morphed}} & \multicolumn{2}{l|}{\textbf{Packed}} \\ 
                                       & \textbf{P}      & \textbf{F1}     & \textbf{P}       & \textbf{F1}       & \textbf{P}       & \textbf{F1}      \\\hline
\textbf{Not Enhanced}                  & 93.4\%          & 87.8\%          & 93.2\%           & 84.7\%            & 39.5\%           & 35.7\%           \\
\textbf{Enhanced}                      & 93.8\%          & 87.7\%          & 93.5\%           & 85.2\%            & 63.5\%           & 51.1\%          \\ \hline
\end{tabular}

\end{table}
It is noteworthy that not all samples were susceptible to morphing or packing. This variability is largely dependent on the individual sample characteristics and the specific malware class. For instance, pre-packed malware typically cannot undergo additional packing using standard UPX. This is the case with the \textit{Elkern} class, in which all the samples are packed with UPX and cannot be packed again.

Our findings demonstrate substantial model resilience against metamorphism but greater susceptibility to packing. This aligns with intuitive expectations and a basic visual inspection of the resulting images. Our meta-engine primarily alters individual instructions, which, in the high-level image representation, does not sufficiently modify the image to induce classifier misclassification. Conversely, packing significantly alters the image patterns and substantially impacts size and dimensions. Consequently, the accuracy of packed malware is considerably lower.
Given these insights, we sought to enhance our malware classifier by implementing a simple technique with negligible overhead increase. By applying packing and morphing to a subset of samples in the original dataset (without introducing new samples), we aimed to train the model to recognize obfuscated samples. This approach effectively serves as a form of artificial data augmentation.
Upon retraining the model with this enhanced dataset and subsequently retesting, we observed improved performance across all datasets, both obfuscated and non-obfuscated, as illustrated in Table~\ref{tab:results_base}. The slight improvement in the non-obfuscated test set can be attributed to the presence of pre-packed samples within the set, not necessarily with the technique we used for the enhancement.

Notably, we observed an approximate 30\% improvement in accuracy on the packed dataset. This substantial enhancement demonstrates the efficacy of this computationally inexpensive technique in mitigating simple obfuscation methods routinely employed by malware authors
\begin{figure}[!htb]
   \begin{minipage}{0.49\textwidth}
     \centering
     \includegraphics[width=.9\linewidth]{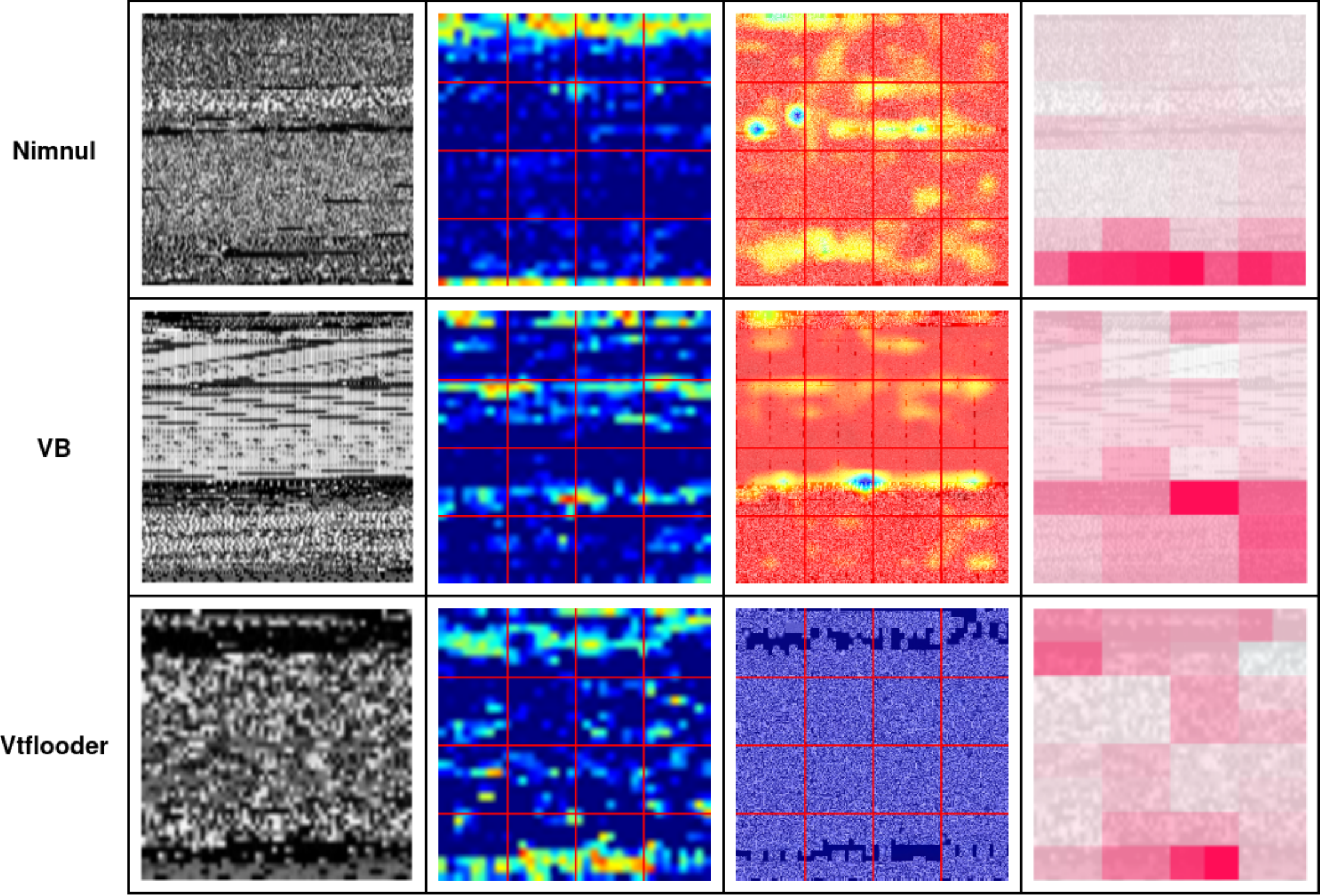}
     \caption{Classes on which explanation techniques identify interesting global patterns.}
    \label{fig:explanations_heuristic}
   \end{minipage}\hfill
   \begin{minipage}{0.49\textwidth}
     \centering
     \includegraphics[width=.9\linewidth]{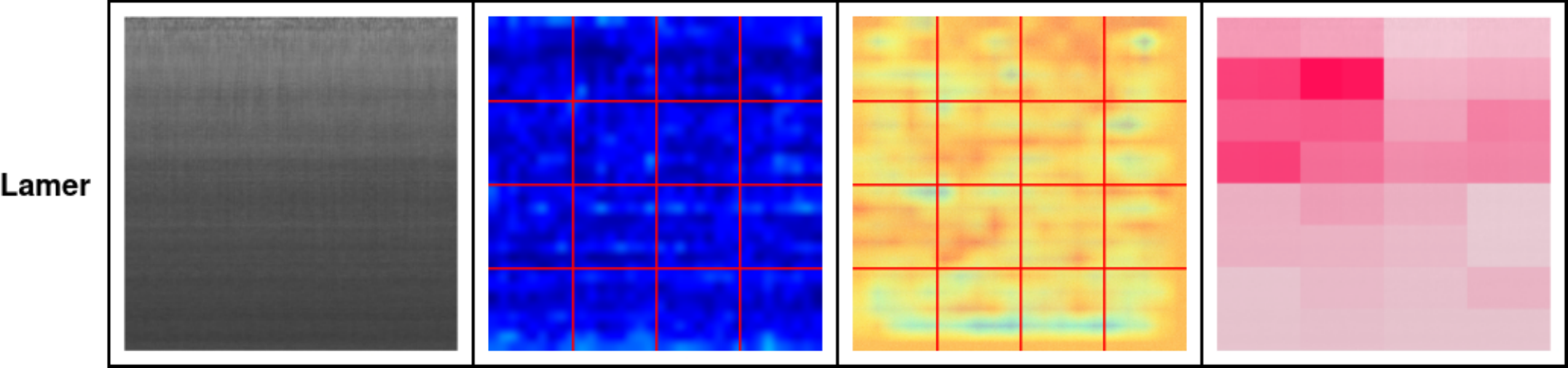}
     \caption{Single class where the explanation techniques cannot identify a pattern.}\label{fig:explanations_random}
   \end{minipage}
\end{figure}

\subsection{Explanations inference}
\label{sec:xai}
In this section, we will examine the explanation maps generated using CAM, SHAP, and occlusion techniques. Our focus will be on identifying instances where the different methods intuitively align in highlighting the key areas of interest for specific class samples. 
To support this analysis, we will also utilize "average sample images", which are created by averaging all the malware images within a class. These visualizations can assist in identifying regions within malware samples where there is minimal intra-class variation (sharper regions), as well as areas with greater intra-class variability (blurred regions). For classes where it is possible, we will also point out artifacts in the heatmaps that may indicate potential family signatures. To better distinguish the coordinates of the image, a grid is provided. 
The first analysis of the explanations has been conducted on the interpretability maps extracted from the model trained on the enhanced training set applied to the basic test set. Each of the results obtained for the families can be categorized in four groups based on the type of inference done on the explanations.
\begin{figure*}
    \centering
    \includegraphics[scale=0.165]{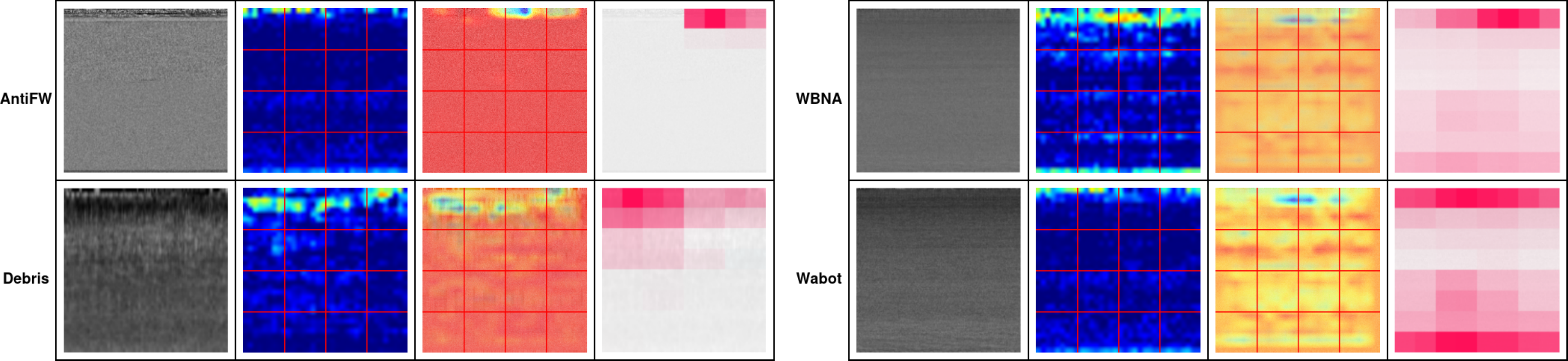}
    \caption{Classes on which explanation techniques generally agree on focused areas of interest.}
    \label{fig:explanations_specific}
\end{figure*}
\subsubsection{Generic patterns}
\begin{figure*}
    \centering
    \includegraphics[scale=0.165]{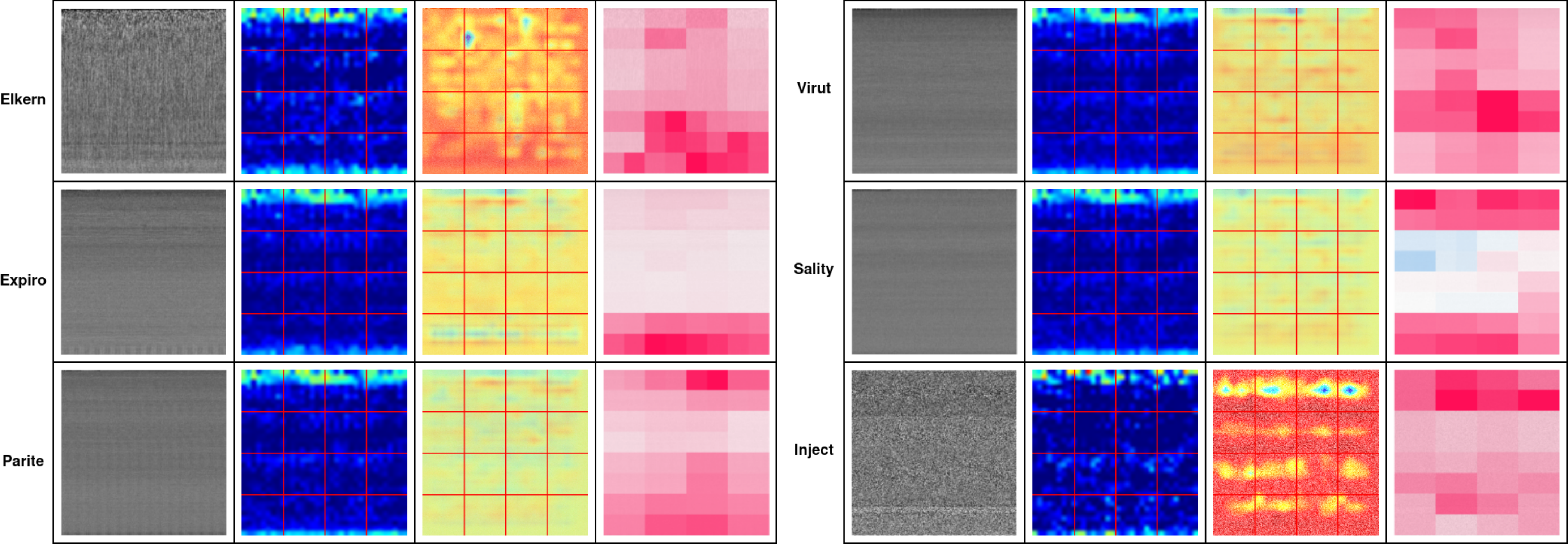}
    \caption{Classes on which explanation techniques result in generic patterns on which is hard to do inference.}
    \label{fig:explanations_generic}
\end{figure*}

The largest group we can identify is composed by the families for which the XAI techniques identify coherent macro patterns used by the CNN to judge their samples. These families are Elkern, Expiro, Parite, Virut, Sality and Inject, seen in Fig.~\ref{fig:explanations_generic}.
For example, for all the families in this group, the CAMs demonstrate a broad focus, highlighting both the top and bottom regions of the image, with different weights based on the family, without pinpointing specific features that appear in areas unique to that class. The SHAP maps obtained for Sality, Inject, Parite and Expiro largely support these findings, in some cases reducing the focus to only one, top or bottom, of the CAM-identified areas. In the case of the occlusion maps, we can identify artifacts in areas similar to CAM in for all the families except Elkern, in which no discernible area seems to appear.
The fact that the CNN focuses on the top parts of the samples is not accidental. Human malware analysts and machine learning classifiers~\cite{manavi2022novel} pay particular attention to the PE header found at the top of PE files because it contains information that can be specific to the sample/ family. Being that the CNN uses these areas, which are known to contain discriminative values useful for humans and other explainable classifiers to analyze malware, we can say with confidence that the CNN is focusing on areas of the files that are objectively interesting from a classification point of view.
\subsubsection{Specific patterns}
The second group identified is composed of AntiFW, Debris, WBNA and Wabot, seen in Fig.~\ref{fig:explanations_specific}. For these families, the XAI techniques identify specific areas used by the CNN to make its predictions. For these classes is particularly interesting the comparison between the XAI maps and the cumulative sample. Starting from AntiFW, The three explainability techniques identify clearly that the PE header is the most important area, which intuitively makes sense being that looking at the cumulative image, the header is the only area seemingly not obfuscated. The CAM identifies the leftmost part as most interesting, while occlusion and SHAP seem to focus on a very specific point in the header just after the area indicated by the CAM. For Debris, the focus is again on the topmost area of the images, but this time not on the PE header but on the generic area right after.  This is important because it indicates that the CNN considers more important the first section just after the PE header than the header itself when classifying Debris samples. Looking at the cumulative image and in the code itself, we can clearly see why. Samples of this family have an empty area just after the PE header that is peculiar to this type of malware. Looking and the other two families in this group, WBNA and Wabot both similar to the ones in previous generic pattern categories, with the difference that they narrow the areas in the header and footer of the images. With WBNA in particular, artifacts are identified by CAM in other areas of the image, but they do not appear in the other XAI techniques.
\subsubsection{Global Peculiar Patterns}
In this group, which includes Nimnul, VB and Vtflooder families, seen in Fig.~\ref{fig:explanations_heuristic}, we can identify very specific patterns in the XAI maps that do not identify a single artifact in their totality are very distinguishable in the areas highlighted than the other maps produced for other families. The cumulative images show a general lack of blurriness, meaning that in the single classes the samples vary only minimally, enabling the CNN to catch on to numerous artifacts that regularly appear in all the samples and uniquely identify the class.
\subsubsection{No discernable pattern}
The last category, fortunately including a single class (Lamer, seen in Fig.~\ref{fig:explanations_random}), represents the situation in which we cannot obtain a clear explanation from the CNN. It is important to consider the fact that the CNN does not perform badly for this class; it just shows that in order to correctly classify samples, it uses different information present in different areas each time it classifies a sample in this class. This shows that the model is solid but is not explainable, at least with the tools used in this research.

\subsubsection{Morphed and packed dataset}
We then moved on studying the changes in the interpretability maps for each class when packing or morphing is applied. In the case of morphing the heatmaps do not significantly change. The reason for this is that the metamorphic changes, however powerful against signature based classifier, do not impact the malware images because their influence is on a very specific part of the whole malware, namely instruction in the code section that can be replaced by equivalent instructions. This type of obfuscation is too specific and sparsely applied that in the generated image do not change the appearing textures, which are what the CNN focuses on~\cite{nataraj2011malware}.
In regards to packing instead the changes are significant. In particular, we noted an emerging trend in the changes of areas where the CNN is focusing on, as demonstrated by particularly CAM and SHAP analysis. The CNN seems to give less weight to the topmost areas of the image and prefers the use of other parts, usually the in the lowermost part of the images. This insight is confirmed by some simple considerations on the inner workings of UPX. This packer, as the majority of other packers, do not touch the \textit{overlay} section and just copy and paste it in the packed samples. This way the end of the sample will be always the same between packed and not packed samples. By using packing and so changing in large part the PE header and sections, the CNN moves it's focus on the only part which remains the same, the \textit{overlay}.

\section{Related works}
\label{sec:related}
In their 2023 paper, Dambra et al..~\cite{dambra2023decoding} investigate state-of-the-art machine learning models to examine their strengths and weaknesses under varying datasets and ground truth conditions. Beyond offering a methodology for addressing these issues, the researchers provide valuable insights into the most significant features utilized by machine learning models and the techniques for their extraction. This paper, along with documents from the Big 2015 project, provides an in-depth analysis of constructing a malware dataset for testing malware classification. It offers various perspectives on the advantages and disadvantages of the different choices researchers face when selecting classes. The major differences have been discussed in Section \ref{sec:dataset}.

Nataraj et al.~\cite{nataraj2011malware} were the first to apply the visualization technique specifically to malware classification by treating malware binaries as gray-scale images. They observe that malware from the same family often share similar visual characteristics. Leveraging this similarity, the document introduces a classification approach using standard image features, namely Gabor filters, eliminating the need for disassembly or code execution. Experiments show promising results, achieving 98\% accuracy in classifying the newly introduced benchmark dataset MalImg, composed of 9,458 samples from 25 different malware families. The authors claim that the technique shows resilience to common obfuscation techniques like section encryption.

In 2022, Vasan et al.~\cite{vasan2020imcec} presented IMCEC, a new technique that integrates the use of different CNNs and traditional machine learning techniques. IMCEC leverages an ensemble of CNNs to detect both packed and unpacked malware. The method relies on the assumption that different CNN architectures provide distinct semantic representations, enabling the extraction of higher-quality features compared to traditional methods. Experimental results demonstrate the effectiveness of IMCEC, achieving over 99\% accuracy for unpacked malware and over 98\% accuracy for packed malware starting from MalImg database. 

In 2023, Chaganti et al.~\cite{chaganti2022image} focus on the use of CNN models to classifying Portable Executable (PE) malware files. Their approach integrates various deep learning model architectures and machine learning classifiers, such as Support Vector Machine (SVM), to evaluate performance across static, dynamic, and image features. The proposed CNN model, using a fusion feature set approach, achieves a 97\% accuracy in classifying malware or benign files. Notably, the model demonstrates robustness and generalizability across different datasets. Visualization techniques are employed to interpret the CNN’s embedding features.

In the same year, Deng et al.~\cite{deng2023mctvd} introduce a new approach to malware visualization using assembly instructions and Markov transfer matrices. The malware classification method, termed MCTVD, is based on three-channel visualization and deep learning. MCTVD generates uniform-sized malware images and utilizes a shallow convolutional neural network architecture. Experimental results demonstrate MCTVD’s effectiveness, achieving a 99.44\% accuracy on the other standard benchmark dataset, Microsoft’s public malware dataset, using 10-fold cross-validation.

In 2024, Kumar et al.~\cite{kumar2024imcnn} presented a model called Intelligent IMCNN. The system involves customizing pre-trained CNN models like VGG16, VGG19, InceptionV3, and ResNet50, trained on the ImageNet database. It integrates feature extraction methods such as principal component analysis (PCA), and singular value decomposition (SVD) are employed, followed by classification using k-nearest neighbor (k-NN), support vector machine (SVM), and random forest (RF) classifiers. The predictive probabilities from these models are combined using a soft voting method for final classification. Evaluation on MalImg datasets and real-world malware samples demonstrates high accuracy (99.36\% and 92.11\% respectively) showing resilience against polymorphic code obfuscation.

A somewhat interesting approach to visualization that follows a similar intuition to what we discovered in our study has been developed by Moreira et al.~\cite{moreira2023improving} in 2023. The basic model that the authors use is a classical CNN, but they apply it to images generated using only the PE header of malware samples. They obtain state-of-the-art results with over 98\% accuracy on a ransomware dataset built for classification. This result corroborates our analysis of the interpretation maps and shows how the PE header (when available) is important for CNNs to classify samples.

Few researchers have directly tackled the problem of explainability applied to black box CNN models. One of the rare ones, specific to the Android platform, is the 2021 paper by Iadarola et al.~\cite{iadarola2021towards} about application of GradCAM~\cite{selvaraju2019gradcam} to a CNN classifier. The paper emphasizes the importance of explainability for analysts to evaluate different models. Experimental results demonstrate the effectiveness of the proposed method, achieving an average accuracy between 96\% and 97\% when evaluating 8446 Android samples across six malware families, and uses interpretability techniques to select the best model among similar performing architectures.

Another research in this direction by the same research group was explored by Ciaramella et al.\cite{ciaramella2022explainable} in 2022. The study focuses on a VGG16-based CNN model and applies GradCAM heatmaps to study the different classes identified by the neural network. They arrive at a similar conclusion as our starting point, in the sense that they identify CAM techniques as a possible avenue to identify visually important artifacts, like the malware payload, in the malware image.

More recently, Galli et al.~\cite{galli2024explainability} used different explainability techniques, among which SHAP to explain neural network malware classification and confronted the results of the different techniques, identifying the one more relevant and relate their results between each other. In their study, the authors use different models and explain them with SHAP, Lime, LRP and attention. They then perturb the samples by using the insights obtained through explainability and observe the difference in performance between the models, highlighting which of the models can be used depending on the situation researchers find themselves in.

Gibert et al.~\cite{gibert2021auditing} in 2021 tackled directly the problem of metamorphic code and how machine learning models perform against it. Their paper addresses the use of metamorphic techniques by malware to evade detection, hindering analysis and thwarting signature-based anti-malware systems. They highlight how, even if deep learning has emerged as the dominant approach for malware detection, limited research on the susceptibility of these models to adversarial attacks has been done. To address this gap, the paper conducts an evaluation of state-of-the-art malware classification methods against common metamorphic attacks. It proposes a novel architecture that improves classification performance by 14.95\% compared to existing approaches. Additionally, the paper explores the use of metamorphic techniques to enhance the training set, demonstrating significant improvements in classifying malware from families with limited samples.
 
An important study on machine learning model performance against packed malware is the 2020 paper "Prevalence and Impact of Low-Entropy Packing Schemes in the Malware Ecosystem" by Mantovani et al.~\cite{mantovani2020prevalence}. Their study mainly focuses on disproving the assumption that a high entropy necessarily means the presence of a packed sample. In the final sections, they study how machine learning classifiers fail to identify packed samples even when supplied with other features in addition to entropy. 

In their 2019 paper, Biondi et al.\cite{biondi2019effective} approach the packing problem by considering not only the effectiveness of packer classifiers but also their efficiency. Through their study, they conclude that a slight loss in accuracy on a small subset of the packed dataset (around 1\%) can cause an improvement in efficiency up to 44 times when using their machine learning model, highlighting the possible choices and tradeoffs available to malware analysts when attacking packed datasets.

Benkraouda et al.\cite{benkraouda2021attack} specifically study how to effectively attack a visualization classifier by using adversarial samples generated with metamorphic-like techniques. Their study demonstrates that even without using complex metamorphic techniques, the adversarial generator is able to fool the classifier, obtaining an evasion rate of 98\% for the samples analyzed.
\section{Conclusion}
\label{sec:conclusion}
While this research has provided valuable insights into the explainability of CNN models using image-based techniques, several avenues for future work remain open. A significant next step in this research would be to bridge the gap between image-based model explanations and the underlying code artifacts that these images represent. Given that the images used in this research are visualizations of binary files, understanding which parts of the code correspond to the most critical regions in the images would allow for a direct mapping between model explanations and source code. The challenge here lies in reversing the abstraction process—going from the visual representation back to the code to identify key code elements. Once these code artifacts (e.g., functions, variables, or patterns) are identified, they could be used as signatures to detect similar patterns in other codebases. This would not only provide a deeper level of interpretability but could also allow for automated code analysis and help with tasks such as vulnerability detection or optimization.
\begin{credits}

\end{credits}
%
%
%
\bibliographystyle{splncs04}
\bibliography{base}

\end{document}